\documentclass{ws-p8-50x6-00}

\def\NPB{{\em Nucl. Phys.} B}

\def\PLB{{\em Phys. Lett.}  B}
\def\PRL{\em Phys. Rev. Lett.}
\def\PRD{{\em Phys. Rev.} D}
\def\PRA{{\em Phys. Rev.} A}
\def\ZPC{{\em Z. Phys.} C}

\begin{document}

\title{Muon \lowercase{$g - 2$} and Electric  Dipole Moments in SUGRA Models}

\author{R. Arnowitt, B. Dutta, B. Hu and Y. Santoso}

\address{
Center For Theoretical Physics, Department of Physics, Texas A$\&$M
University, College Station TX 77843-4242, USA\\ 
E-mail: arnowitt@physics.tamu.edu}

\maketitle

\abstracts{
The SUSY contribution to the muon magnetic moment anomaly, $a_{\mu}^{\rm
SUGRA}$, and 
the electron electric dipole moment, $d_e$, is discussed within the framework 
of a modified mSUGRA model where the magnitudes of the soft breaking masses 
are universal, but arbitrary phases are allowed. It is shown analytically 
how the cancellation mechanism can allow for large phases (i.e. $\theta_B
\stackrel{<}{\sim} 0.4$) and still suppress the value of $d_e$ below its
current experimental 
bound. The dependence of $a_{\mu}^{\rm SUGRA}$ on the CP violating phases are 
analytically examined, and seen to decrease it but by at most a factor 
of  about two. This reduction would then decrease the upper bound on $m_{1/2}$ 
due to the lower bound of Brookhaven data, and hence lower the SUSY mass 
spectrum, making it more accessible to accelerators. At the electroweak 
scale, the phases have to be specified to within a few percent to satisfy 
the experimental bound on $d_e$, but at the GUT scale, fine tuning below 1\% is 
required for lower values of $m_{1/2}$. This fine tuning problem will become 
more serious if the bound on $d_e$ is decreased.
}

\section{Introduction}

In supersymmetry, the interaction of charginos ($\tilde{\chi}^{\pm}_i$,
$i=1,2$), neutralinos 
($\tilde{\chi}^0_j$, $j=1,2,3,4$) and sleptons ($\tilde{e}_k$,
$\tilde{\mu}_k$, $k=1,2$;
$\tilde{\nu}_e$, $\tilde{\nu}_{\mu}$) with the 
leptons ($l_m = e, \mu, \tau$) gives rise to electromagnetic vertices $l_m -
l_n - \gamma$. Thus the basic diagrams for the diagonal muon 
interaction shown in Fig. 1 gives rise to anomalous contributions to the 
muon magnetic moment $a_{\mu} = (g - 2)/2$, and could possibly produce an
electric 
dipole moment, $d_{\mu}$. Similar diagrams with $\mu \rightarrow e$ and
$\tilde{\nu}_{\mu}, \tilde{\mu} \rightarrow \tilde{\nu}_e, \tilde{e}$ can give
rise to $a_e$ and $d_e$, while the off diagonal diagram could allow for the 
decay $\mu \rightarrow e + \gamma$. The different possibilities, however,
involve 
different physics. Corrections to the anomalous magnetic moments of the 
leptons are always present in supersymmetry, and the recent results of the 
Brookhaven E821 experiment~\cite{b1} showing a $2.6\; \sigma$ deviation from
the Standard Model prediction,
\begin{equation}
   a_{\mu}^{\rm exp} - a_{\mu}^{\rm SM} = 43 (16) \times 10^{-10}     
   \label{eq1}
\end{equation}
shows the possibility of a SUSY contribution. Indeed, in the initial 
calculations~\cite{b2,b3} based on supergravity grand unification models
(mSUGRA~\cite{b4}), it was predicted~\cite{b3} that a deviation should show up
when the 
experimental sensitivity reached that of the Brookhaven experiment. 
However, for electric dipole moments to be non-zero requires in addition 
the presence of CP violating phases. The current experimental bounds on $d_e$ 
is~\cite{b5}
\begin{equation}
      d_e < 4.3 \times 10^{-27} 	\label{eq2}
\end{equation}
and this bound is expected to be reduced by a factor of 2-3 in the near 
future~\cite{b6}.  

\begin{figure}[htb]
\begin{center}
\setlength{\unitlength}{0.015in}
\begin{picture}(240,65)(0,0)
\put(0,56){\line(1,0){30}}
\put(30,56){\line(1,0){40}}
\put(70,56){\line(1,0){30}}
\put(140,56){\line(1,0){30}}
\put(170,56){\line(1,0){40}}
\put(210,56){\line(1,0){30}}
\qbezier(30,56)(50,16)(70,56)
\qbezier(170,56)(190,16)(210,56)
\qbezier(50,28)(55,26)(50,24)
\qbezier(50,24)(45,22)(50,20)
\qbezier(50,20)(55,18)(50,16)
\qbezier(50,16)(45,14)(50,12)
\qbezier(50,12)(55,10)(50,8)
\qbezier(50,8)(45,6)(50,4)
\qbezier(50,4)(55,2)(50,0)
\qbezier(190,28)(195,26)(190,24)
\qbezier(190,24)(185,22)(190,20)
\qbezier(190,20)(195,18)(190,16)
\qbezier(190,16)(185,14)(190,12)
\qbezier(190,12)(195,10)(190,8)
\qbezier(190,8)(185,6)(190,4)
\qbezier(190,4)(195,2)(190,0)
\put(30,56){\circle*{2}}
\put(70,56){\circle*{2}}
\put(170,56){\circle*{2}}
\put(210,56){\circle*{2}}
\put(5,48){$\mu$}
\put(93,48){$\mu$}
\put(145,48){$\mu$}
\put(233,48){$\mu$}
\put(55,6){$\gamma$}
\put(195,6){$\gamma$}
\put(50,61){$\tilde{\nu}_{\mu}$}
\put(26,35){$\tilde{\chi}_i^{\pm}$}
\put(190,62){$\tilde{\chi}_j^{0}$}
\put(170,36){$\tilde{\mu}$}
\end{picture}
\end{center}
\caption{Diagrams contributing to $a_{\mu}$ and $d_{\mu}$ involving
intermediate 
chargino-sneutrino states and intermediate neutralino-smuon states.
\label{fig1}}
\end{figure}
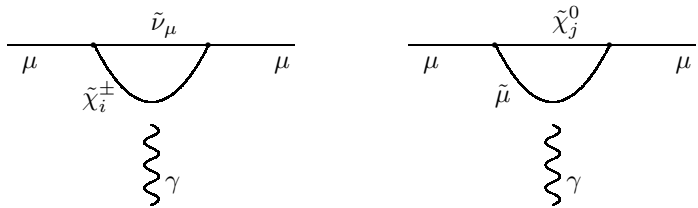

In this paper, we examine the magnetic moment and electric dipole moment 
phenomena within the framework of gravity mediated SUGRA models, and assume 
that the Brookhaven anomaly is real~\cite{b7} and due to SUSY. We will assume a
$2\;\sigma$ range for $a_{\mu}^{\rm SUGRA}$
\begin{equation}
  11 \times 10^{-10}  <  a_{\mu}^{\rm SUGRA} < 75 \times 10^{-10}   \label{eq3}
\end{equation}
 If we assume universal soft 
breaking in the first two generation, then $m_{\tilde{\mu}_k} =
m_{\tilde{e}_k}$, and $m_{\tilde{\nu}_{\mu}} = 
m_{\tilde{\nu}_e}$. Then both $a_{\mu}$ and $d_e$ can be obtained from the same
complex amplitude $A$:
\begin{equation}
    a_{\mu}^{\rm SUGRA}  = - \frac{\alpha}{4 \pi \sin^2 \theta_W} \, m_{\mu}^2
    \, {\rm Re}[A]                      \label{eq4}    
\end{equation}
\begin{equation}
    d_e = - \frac{\alpha}{8\pi \sin^2 \theta_W} \, m_e \, {\rm Im}[A]  
    \label{eq5} 
\end{equation}
There has been a great deal of recent analysis of $a_{\mu}$ within the framework 
of mSUGRA models~\cite{b8,b9}  
showing that indeed for current theory constrained by all accelerator and
non-accelerator bounds, 
$a_{\mu}^{\rm SUGRA}$ can be expected to lie within the range of Eq. (3). It is
thus at first sight puzzling that $d_e$ is so small, since it arises from the
same amplitude $A$ that gives rise to a large $a_{\mu}^{\rm SUGRA}$. Two
possible 
explanations for this are the following: (1) The CP violating phases 
appearing in $A$ are anomalously small, and in fact it is possible to build 
reasonably natural models where this can happen~\cite{b10}. (2) The CP
violating phases are indeed $O(1)$ (as the CKM phase is), but there are
cancelations between 
the two diagrams of Fig. 1 suppressing the value of $d_e$~\cite{b11}. This
possibility 
appears more preferable, and there has been considerable analysis within 
that framework~\cite{b12} and we will consider it here. However, 
this leads immediately to the following question: if cancellations occur in 
$d_e$, why do corresponding cancellations not occur also in $a_{\mu}^{\rm
SUGRA}$\, ? In 
the following we will answer this question and show analytically how one 
may have large phases such that $d_e$ is suppressed to nearly zero, but 
$a_{\mu}^{\rm SUGRA}$ is reduced by less than a factor of 2 (so that agreement
with 
the Brookhaven data is maintained). The remaining question then is whether 
fine tuning of the phases is needed to suppress $d_e$ to the level of 
Eq. (2) when the phases are large. We will see that significant fine tuning has 
started to occur at the GUT scale, and this problem will become more 
serious if the upper bound on $d_e$ is reduced further.

\section{SUGRA Models}

We consider here a generalization of the usual mSUGRA model allowing for CP 
violating phases. At the GUT scale $M_G$, the theory depends upon the 
following parameters: $m_0$, the universal scalar soft breaking mass; $m_i = 
|m_{1/2}| e^{i\phi_i}$, $i = 1,2,3$ the three gaugino masses; $A_0 = 
|A_0| e^{i \alpha_0}$, the cubic soft breaking mass; $B_0 = |B_0| e^{i
\theta_{B_0}}$, 
the quadratic soft breaking mass; and $\mu_0 = |\mu_0| e^{i \theta_{\mu}}$,
the 
Higgs mixing mass. One is always free to set one of the gaugino phases to 
zero and we chose $\phi_2 = 0$. Radiative breaking of $SU(2) \times U(1)$ at
the 
electroweak scale determines  $\theta_{\mu}$ according to (with the convenient
choice that the Higgs VEVs be real)
$\theta_{\mu} = - \theta_B + f_1(-\theta_B + \alpha_l,  -\theta_B + \alpha_q)$  
where $f_1$ is a loop correction, $\alpha_{(l,q)}$ are the (lepton, quark)
phases of $A$ at the electroweak 
scale, and $\theta_B$ is the $B$ phase at the electroweak scale. In addition 
$|\mu|^2$ and $|B|$   are determined by the usual formulae.  Thus the theory is 
defined by four real parameters, $m_0$, $|m_{1/2}|$, $|A_0|$, and $\tan\beta =
\langle H_2 \rangle / \langle H_1 \rangle$, 
and four new CP violating phases: $\theta_{B_0}$, $\phi_1$, $\phi_3$,
$\alpha_0$. 
However, since we are here examining only the electron EDM, our results 
depend only weakly on $\phi_3$ and $\alpha_0$. Thus the two important phases
are $\theta_{B_0}$ and $\phi_1$.

In carrying out the calculations discussed below, it is important to impose 
all the known accelerator and non-accelerator bounds on the SUSY parameter 
space (including coannihilation effects in $\Omega_{\tilde{\chi}_1^0} h^2$),
and these bounds need to be calculated accurately. For a discussion of these see
Ref. 13. Finally we mention that we scan 
the parameter space over the range $m_0, |m_{1/2}| < 1$ TeV, $|A_0/m_{1/2}| <
4$, and $2 < \tan\beta < 50$.

\section{Suppression of \lowercase{$d_e$}}

It was realized from the beginning that $a_{\mu}$ is an increasing function of 
$\tan\beta$~\cite{b2,b3}. The Brookhaven data favors $\tan\beta >
5-7$~\cite{b8}, and larger 
$\tan\beta$ is consistent with the data.  In order to see analytically the 
effect of the CP violating phases then, we consider the leading part of the 
complex amplitude for large $\tan\beta$ calculated in Ref. 13. 
Using Eq. (4), one can write $a_{\mu}^{\rm SUGRA}$ in the form
\begin{equation}
               a_{\mu}^{\rm SUGRA}  =  A \cos \theta_{\mu} + B \cos(\theta_{\mu}
	       + \phi_1) + b 
\cos\theta_{\mu} + c       \label{eq6}
\end{equation}
 $A$ comes 
from the chargino diagram, while $B$, $b$ and $c$ arise from the neutralino 
diagram. In general, the chargino contributions are largest  and the 
parameter $a = B/(A + b)$ is $\cong 0.10 - 0.45$. ($b$ and $c$ are generally
small). 
Similarly, Eq.(5) for $d_e$ has the general form
\begin{equation}
                \frac{d_e}{e} = - \frac{m_e}{2 m_{\mu}^2} (A +b)\, [
		\sin\theta_{\mu} + a \sin(\theta_{\mu} + 
\phi_1)]                       \label{eq7}
\end{equation}
One sees that a priori, the amplitudes for $a_{\mu}^{\rm SUGRA}$ and $d_e$ are
of large 
size, being scaled by the largest amplitudes $A + b$. However, one can 
suppress $d_e$, and in fact even obtain $d_e = 0$ if the phases obey
$\sin \theta_{\mu} + a \sin(\theta_{\mu} + \phi_1) = 0$, 
or alternately, using  Eq. (6) (neglecting the small loop corrections),
if~\cite{b18}
\begin{equation}
       \tan\beta(\theta_B)  = \frac{a \sin \phi_1}{ (1 + a 
\cos \phi_1)}        \label{eq8}    
\end{equation}
Since $a$ is not small, one sees that $\theta_B$ need not be small to
accommodate 
even $d_e = 0$. This is essentially the origin of the cancellation
effect~\cite{b11} 
for large SUSY CP violating phases.

One may now insert Eq. (8) into Eq. (6) to see the effect the phases have 
on $a_{\mu}^{\rm SUGRA}$:
\begin{equation}
           a_{\mu}^{\rm SUGRA}(\theta_B, \phi_1) = a_{\mu}^{\rm SUGRA}(0,0)
	   \frac{\cos
	   \theta_B}{|\cos \theta_B|} Q(\phi_1)      \label{eq9}
\end{equation}
where
\begin{equation}
           Q(\phi_1) = \frac{[1 + 2 a \cos \phi_1 + a^2]^{1/2}}{(1 + a)}; \qquad
	   0.5 \stackrel{<}{\sim} Q \leq 1   \label{eq10}      
\end{equation}
Thus the effect of the non-zero phases which reduce $d_e$ to zero is to reduce 
the magnitude of $a_{\mu}^{\rm SUGRA}$ by at most a factor of two. Further,
since 
experimentally $a_{\mu}^{\rm SUGRA} > 0$, one requires $\cos \theta_B > 0$ and
so
\begin{equation}
                      \theta_B > 0 \quad {\rm for} \quad  0 < \phi_1 < \pi; 
		      \qquad   \theta_B < 0 \quad {\rm for} \quad  \pi 
<  \phi_1 < 2\pi       \label{eq11}
\end{equation}
and so as $\phi_1$ varies over the entire range, one has $|\theta_B|
\stackrel{<}{\sim} 0.4$.

In summary then, even if $d_e$ were zero, the cancellation mechanism~\cite{b11}
can 
accommodate large phases ($|\theta_B| \stackrel{<}{\sim} 0.4$), and give
acceptable predictions 
for the muon magnetic moment anomaly~\cite{b18}. The question, however, is:
given 
the current bounds Eq.(2) on $d_e$, can one have large CP violating phases 
without unreasonable fine tuning of $\theta_B$ or $\phi_1$? We turn to this 
question next.

\section{Large Phases and Fine Tuning}

Since the experimental upper bound on $d_e$ is so small, one might expect that 
when the SUSY CP violating phases are large, fine tuning might be required 
to obtain the necessary amount of suppression. In order to quantify this, 
we define for any phase $\phi$ the quantity
\begin{equation}
      R(\phi) = \frac{(\phi_1 - \phi_2)}{(\phi_1 + \phi_2)/2}        
      \label{eq12}      
\end{equation}
where $\phi_1$ and $\phi_2$ are the largest and smallest values of $\phi$ that 
satisfy the bound of Eq. (2). Thus $R(\phi)$ is a measure of how tightly 
constrained a phase must be to satisfy the experimental bounds. How much 
fine tuning one can tolerate is, of course, a matter of individual taste. 
However, we will take here as a benchmark the requirement that for any 
phase, $R(\phi) > 0.01$.

For a grand unified model, presumably parameters at the GUT scale are the 
more fundamental ones, and they are to be determined by some higher theory. 
Thus fine tuning at $M_G$ can represent a serious problem.  The sensitive
parameter in this case is 
$\theta_{B_0}$ (the $B$ phase at $M_G$). To understand analytically what may be 
occurring, we consider the RGE for the low and intermediate $\tan\beta$ region
where an analytic expression exists. 
One finds~\cite{b10}
\begin{equation}
   B = B_0 - \frac{1}{2}(1 - D_0)A_0 - \Sigma \Phi_i \,  |m_{1/2}| \,
e^{i \phi_i}         \label{eq13}
\end{equation}
where $D_0 = 1 - m_t^2/(200 \sin \beta)^2 \stackrel{<}{\sim} 0.25$ and $\Phi_i =
O(1)$. Taking the 
imaginary part one finds
\begin{equation}
    |B| \sin \theta_B = |B_0| \sin \theta_{B_0} - \frac{1}{2}(1 - D_0) |A_0| 
\sin \alpha_0 - \Sigma \Phi_i \, m_{1/2} \, \sin \phi_i     \label{eq14}
\end{equation}
For fixed $\alpha_0$ and $\phi_i$, one can relate the range of $\theta_B$,
$\Delta \theta_B$, allowed by 
radiative electroweak breaking in terms of the range of $\theta_{B_0}$, $\Delta
\theta_{B_0}$:
\begin{equation}
    \Delta(\theta_{B_0})  \cong  \frac{|B|}{|B_0|} 
    \Delta(\theta_B) 
\label{eq15}
\end{equation}
However, radiative breaking at the electroweak scale shows that $|B|$ gets 
small as $\tan \beta$ grows i. e. $|B| = (1/2) \sin 2 \beta(m_3^2/|\mu|)$. Hence
we expect that
\begin{equation}
     \Delta \theta_{B_0} \ll \Delta \theta_B 
\label{eq16}
\end{equation}
and fine tuning may occur at the GUT scale. An example of what happens is 
shown in Fig.3 where $R(\theta_{B_0})$ is plotted as a function of $m_{1/2}$
for 
$\tan\beta = 40$, $A_0 = 0$ for $\phi_1= 1.6$, 1.2, 0.9, 2.3, and 2.6. One sees
that 
for a wide range of $\phi_1$, $R(\theta_{B_0})$ falls below  0.01, and if one
were to 
impose the condition that $R > 0.01$, it would eliminate a significant 
portion of the low $m_{1/2}$ parameter space. In the near future, one may
expect 
the bounds on $d_e$ to be reduced by a factor of 2 to 3~\cite{b6}. This would 
exacerbate the fine tuning problem at the GUT scale.

\smallskip
\begin{figure}[htb]
\begin{center}
\epsfxsize=20pc 
\epsfbox{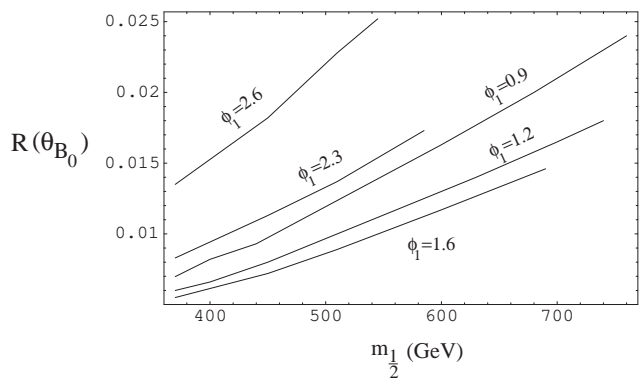}
\end{center}
\caption { 
$R(\theta_{B_0})$ as a function of $m_{1/2}$ for $\tan \beta = 40$, $A_0
= 0$ for (from 
bottom to top) $\phi_1 = 1.6$, 1.2, 0.9, 2.3 and 2.6. $^{13}$  \label{fig4}}
\end{figure}

\section{Conclusion}

If the SUSY CP violating phases are small, i.e. $O(10^{-2})$, then the bounds 
on $d_e$ can be satisfied, and the effects of $d_e$ disconnect from $g_{\mu} - 2$.
The 
Brookhaven E821 experiment, plus other experimental constraints, then leads 
to the following results for mSUGRA models~\cite{b8}: The light Higgs, $h$,
mass 
bound combined with the $b \rightarrow s+\gamma$ constraint gives rise to a
lower bound 
on $m_{1/2}$ of $m_{1/2} \stackrel{>}{\sim} 300-400$ GeV. The lower bound on
the $a_{\mu}$ anomaly then 
produces an upper bound on $m_{1/2}$ of $m_{1/2} < 585 (845)$ GeV at the 90\%
(95\%) 
C.L. for $\tan \beta \leq 50$. For mSUGRA, $a_{\mu}^{\rm SUGRA}$ is bounded
from above with $a_{\mu}^{\rm
SUGRA} \stackrel{<}{\sim} 50 \times 10^{-10}$. One can then predict the mSUGRA
discovery reaches for 
accelerators. Thus at the 90\% C.L. one finds that the Tevatron should see 
the light Higgs only, while a 500 GeV NLC would be able to see only $h$, the 
light stop squark and perhaps the light selectron. (One would need a higher 
energy LC to see more of the SUSY spectrum.) The LHC, of course, could see 
the entire SUSY spectrum. In addition, the lower bounds of dark matter 
detection rates would be raised, since the upper bound on $m_{1/2}$ has been 
lowered, and $\mu > 0$.

If the CP violating phases are large, they effect both $d_e$ and $a_{\mu}$. The 
experimental bound on $d_e$ can then still be satisfied if the cancelation 
mechanism~\cite{b11} occurs. One finds then the following~\cite{b18}: The
cancelations 
can be understood analytically [Eq. (11)] and are seen to lead to $\theta_B$ as 
large as $\sim 0.4$ with large gaugino phase $\phi_1$. The value of
$a_{\mu}^{\rm SUGRA}$ is 
reduced by at most a factor of about two, and so agreement with the 
Brookhaven $a_{\mu}$ data can still be satisfied. (The reduction of
$a_{\mu}^{\rm SUGRA}$ 
will, however, lower the upper bound on $m_{1/2}$ and thus lower the SUSY mass 
spectrum, increasing the reach of accelerators.)  The experimental bounds 
on $d_e$ can generally be satisfied with a fine tuning of phases of a few 
percent at the electroweak scale. However at $M_G$, fine tuning $< 1\%$ is
needed 
for $\theta_{B_0}$ in the lower $m_{1/2}$ region. This fine tuning will become
more 
serious if the experimental bound on $d_e$ is lowered further.

\section*{Acknowledgments}
This work was supported in part by a National Science Foundation grant 
number PHY-0070964.

\end{document}